\documentclass[12pt,preprint]{emulateapj}

\usepackage{amsmath}
\usepackage{bm}

\slugcomment{\today}

\shorttitle{Supernova Relic Neutrino and Metallicity Evolution}
\shortauthors{Nakazato et al.}

\begin{document}


\title{Spectrum of the Supernova Relic Neutrino Background and \\ Metallicity Evolution of Galaxies}


\author{Ken'ichiro Nakazato\altaffilmark{1}, Eri Mochida\altaffilmark{1}, Yuu Niino\altaffilmark{2}, and Hideyuki Suzuki\altaffilmark{1}}

\email{nakazato@rs.tus.ac.jp}


\altaffiltext{1}{Department of Physics, Faculty of Science \& Technology, Tokyo University of Science, 2641 Yamazaki, Noda, Chiba 278-8510, Japan}
\altaffiltext{2}{Division of Optical \& Infrared Astronomy, National Astronomical Observatory of Japan, 2-21-1 Osawa, Mitaka, Tokyo 181-8588, Japan}


\begin{abstract}
The spectrum of the supernova relic neutrino (SRN) background from past stellar collapses including black hole formation (failed supernovae) is calculated. The redshift dependence of the black hole formation rate is considered on the basis of the metallicity evolution of galaxies. Assuming the mass and metallicity ranges of failed supernova progenitors, their contribution to SRNs is quantitatively estimated for the first time. Using this model, the dependences of SRNs on the cosmic star formation rate density, shock revival time and equation of state are investigated. The shock revival time is introduced as a parameter that should depend on the still unknown explosion mechanism of core collapse supernovae. The dependence on equation of state is considered for failed supernovae, whose collapse dynamics and neutrino emission are certainly affected. It is found that the low-energy spectrum of SRNs is mainly determined by the cosmic star formation rate density. These low-energy events will be observed in the Super-Kamiokande experiment with gadolinium-loaded water.
\end{abstract}





\section{Introduction}\label{sec:intro}

Since the creation of the Universe, many generations of stars have been born and died. During the cosmic evolution, stars eject synthesized elements by stellar winds or explosions such as supernovae, and the ejecta are mixed with the interstellar gas. Therefore, the mass fraction of elements heavier than carbon (metallicity), $Z$, increases gradually with the cosmic time. Meanwhile, many neutrinos are emitted from core collapse supernova (CCSN) explosions of massive stars and accumulate to give a diffuse background radiation that is redshifted owing to cosmic expansion. These neutrinos are called the supernova relic neutrino (SRN) background, or the diffuse supernova neutrino background (DSNB) in some papers.

\begin{figure}
\plotone{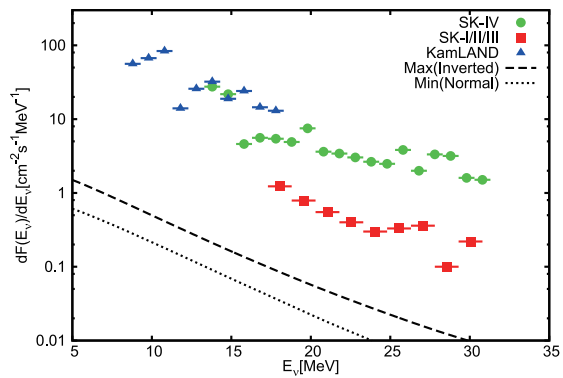}
\caption{90\% C.L. differential upper limits on $\bar\nu_e$ flux of SRNs. The squares, circles and triangles are results for Super-Kamiokande \citep[SK-I/II/III,][]{bays12}, Super-Kamiokande with a neutron-tagging \citep[SK-IV,][]{sk4srn} and KamLAND \citep[][]{kamsrn}. Dashed and dotted lines correspond to our theoretical models with maximum and minimum values of SRN event rate, respectively (see also Table~\ref{tab:enste}).}
\label{fig:pre}
\end{figure}

Neutrinos emitted from a supernova have actually been detected for SN1987A \citep[e.g.,][]{hirata87,bionta87,alex88}. In the observation of SRNs, on the other hand, terrestrial neutrino detectors are affected by various backgrounds such as solar neutrinos, reactor neutrinos, atmospheric neutrinos and contamination by cosmic muon events, radio activity events and so forth. However, some observational upper bounds for the flux of SRNs have been reported \citep[e.g.,][]{malek03}. Roughly speaking, all species of neutrinos ($\nu_{e}$, $\bar{\nu}_{e}$, $\nu_{\mu}$, $\bar{\nu}_{\mu}$, $\nu_{\tau}$, $\bar{\nu}_{\tau}$) are equally emitted from a supernova with average energies of $\sim$10~MeV. Nowadays, SRNs with $\bar{\nu}_{e}$ of approximately 20~MeV are expected to be observable in running experiments. The most stringent limits reported for $\bar{\nu}_{e}$ flux were obtained in the Super-Kamiokande experiment as $<$0.1-1~cm$^{-2}$~s$^{-1}$~MeV$^{-1}$ for neutrino energies between 17.3~MeV and 30.8~MeV \citep[][]{bays12} and in the KamLAND experiment as $<$10-100~cm$^{-2}$~s$^{-1}$~MeV$^{-1}$ between 8.3~MeV and 18.3~MeV \citep[][]{kamsrn}. Super-Kamiokande derived a new upper limit of $<$5-30~cm$^{-2}$~s$^{-1}$~MeV$^{-1}$ for energies between 13.3~MeV and 17.3~MeV by performing a new analysis with a neutron-tagging technique \citep[][]{sk4srn}. In Figure~\ref{fig:pre}, we show the upper limits for $\bar{\nu}_{e}$ flux with theoretical estimations presented later in this paper. For $\nu_{e}$ flux, the SNO experiment obtained an upper limit of 70~cm$^{-2}$~s$^{-1}$~MeV$^{-1}$ for energies between 22.9~MeV and 36.9~MeV \citep[][]{snosrn}. These observational upper limits are larger than various theoretical predictions \citep[e.g.,][and references therein]{ando04,beacom10}. Nevertheless, the Super-Kamiokande upper limit is reasonably close to the predictions; thus, it is expected that SRNs will be observed in the near future.

Cosmic metallicity evolution has been proven by observations of galaxies \citep[e.g.,][hereafter M08]{m08}. Recently, the correlation between the metallicity and the star formation rate (SFR) of galaxies has been studied for various ranges of the cosmic redshift parameter $z$ \citep[e.g.,][]{mannu10,yabe12,yabe14,zahid14}. Since metals have strong effects on opacity and control the cooling rate and luminosity of star-forming clouds, SFR depends on the metallicity. On the other hand, SFR determines the amount of metals ejected from stars and governs the metallicity evolution. Among the newly born stars, those with mass of $\gtrsim \! 10M_{\odot}$ undergo gravitational collapse to cause a CCSN or, if they fail to explode, form a black hole. Combining the SFR of individual galaxies with the galaxy distribution, the cosmic star formation rate density (CSFRD) and total core-collapse rate can be determined at each redshift. Therefore, the flux of SRNs is closely related to the evolution of galaxies \citep[][]{totani96}.

Metallicity affects not only star formation but also stellar evolution. Since the stellar opacity and mass loss rate vary with the metallicity, the final fate of stars depends on the metallicity as well as the initial mass, $M$. Some progenitors may result in black-hole-forming failed supernovae depending on their mass and metallicity. They will contribute to the flux of SRNs because neutrinos are also emitted from them with higher luminosity and mean energy than those from ordinary CCSN \citep[e.g.,][]{sumi06}. The spectrum of SRNs that will be observed is the convolution of the neutrino spectra emitted from core collapses of progenitors with various masses and metallicities. Conversely, the observation of SRNs will provide useful information not only on supernova neutrinos themselves but also on cosmological evolutions of the star formation and metallicity.

In this study, we construct a model spectrum of SRNs using the metallicity distribution function derived from models of galaxy evolution for the first time. Using this model, we investigate some uncertainties in SRNs, namely CSFRD, the explosion mechanism and the nuclear equation of state (EOS). It is pointed out that most of the theoretical models for galaxy formation underpredict CSFRD compared with observations at $z \gtrsim 1$ \citep[e.g.,][hereafter K13]{koba13}. The explosion mechanism of a CCSN is still uncertain. Most numerical simulations reveal that the shock wave launched at the bounce of the inner core stalls on the way to the core surface and that some mechanism such as neutrino heating acts to revive the shock wave leading to the supernova explosion \citep[e.g.,][]{kotake12}. The nuclear EOS, which is also still unknown, affects the collapse dynamics and neutrino emission, especially for failed supernovae \citep[e.g.,][]{sumi06}.

This paper is organized as follows. In \S~\ref{sec:srn}, we present the formulation of the SRN spectrum. Issues on neutrino oscillation are also given. In \S~\ref{sec:gal}, the models of galaxy evolution used in this study are described. We explain numerical models of the neutrino spectra emitted from various progenitors in \S~\ref{sec:snn}. Here we introduce the shock revival time as a parameter that reflects the unknown explosion mechanism. In addition, we report a new result for a failed supernova with a different EOS. In \S~\ref{sec:event}, we show the results for the spectrum of SRNs and the event rate for Super-Kamiokande over 1 year. We also investigate the uncertainties of the CSFRD, shock revival time and EOS for black hole formation. Finally, \S~\ref{sec:concl} is devoted to a conclusion and discussion.

\section{Formulation of Supernova Relic Neutrino Background}\label{sec:srn}

In this study, we construct a model spectrum of SRNs taking into account cosmic metallicity evolution. Since SRNs originate from various progenitors, their dependence of the neutrino emission affects the spectrum. Here we assume that the progenitors are characterized by their initial mass, $M$, and metallicity, $Z$. Then the flux of SRNs on the Earth is written as
\begin{eqnarray}
\frac{{\rm d}F(E_\nu)}{{\rm d}E_\nu} & = & c \int_0^{z_{\rm max}} \frac{{\rm d}z}{H_0 \sqrt{\Omega_m (1+z)^3 + \Omega_\Lambda}} \times \nonumber \\
 & & \Biggl[ R_{\rm CC} (z) \int^{Z_{\rm max}}_{0} \psi_{\rm ZF} (z, Z)  \nonumber \\
 & & \biggl\{ \int^{M_{\rm max}}_{M_{\rm min}} \psi_{\rm IMF} (M) \frac{{\rm d}N(M, Z, E^\prime_\nu)}{{\rm d}E^\prime_\nu} \,{\rm d}M \biggr\} \,{\rm d}Z \Biggr], \nonumber \\
 & &
\label{eq:srn}
\end{eqnarray}
with velocity of light $c$ and cosmological constants $H_0=70~{\rm km}~{\rm s}^{-1} \,{\rm Mpc}^{-1}$, $\Omega_m=0.3$ and $\Omega_\Lambda=0.7$. The neutrino energy on the Earth, $E_\nu$, is related to that at the redshift $z$, $E^\prime_\nu$, as $E^\prime_\nu=(1+z)E_\nu$. The total core-collapse rate, $R_{\rm CC} (z)$, is determined by CSFRD, which is stated in \S~\ref{sec:csfrd}. The neutrino number spectrum from the core collapse of a progenitor with mass $M$ and metallicity $Z$, ${\rm d}N(M, Z, E^\prime_\nu)/{\rm d}E^\prime_\nu$, includes the effect of neutrino oscillation, which is stated below. Note that, in our model, not only ordinary supernovae but also black-hole-forming collapses without an explosion are considered. Meanwhile, $\psi_{\rm IMF} (M)$ and $\psi_{\rm ZF} (z, Z)$ are the initial mass function and metallicity distribution function of progenitors, respectively, which are normalized as $\int^{M_{\rm max}}_{M_{\rm min}} \psi_{\rm IMF} (M) {\rm d}M=\int^{Z_{\rm max}}_{0} \psi_{\rm ZF} (z, Z) {\rm d}Z=1$. While the initial-mass dependence of progenitors was studied previously \citep[][]{self06,luna12,self13c}, the metallicity dependence is also considered in our study. Note that $\psi_{\rm ZF} (z, Z)$ is a function of the redshift $z$ due to the cosmic metallicity evolution. We use the metallicity distribution function derived from models of galaxy evolution later in \S~\ref{sec:metal}.

As already mentioned, we should take into account neutrino oscillation in the evaluation of SRN spectra. Hereafter, with the Super-Kamiokande experiment in mind, we concentrate on $\bar{\nu}_e$. The number spectrum of $\bar{\nu}_e$ in Equation~(\ref{eq:srn}), ${\rm d}N_{\bar{\nu}_e}(M, Z, E_\nu)/{\rm d}E_\nu$, is a mixture of spectra of neutrinos originally produced as $\bar{\nu}_e$, $\bar{\nu}_{\mu}$, and $\bar{\nu}_{\tau}$ and is written as
\begin{eqnarray}
 & & \frac{{\rm d}N_{\bar{\nu}_e}(M, Z, E_\nu)}{{\rm d}E_{\nu}} \nonumber \\
 & & = \bar{P}_{ee}\,\frac{{\rm d}N_{\bar{\nu}_e}^{0}(M, Z, E_\nu)}{{\rm d}E_{\nu}} + \bar{P}_{\mu e}\,\frac{{\rm d}N_{\bar{\nu}_\mu}^{0}(M, Z, E_\nu)}{{\rm d}E_{\nu}} \nonumber \\
 & & \quad {}+ \bar{P}_{\tau e}\,\frac{{\rm d}N_{\bar{\nu}_\tau}^{0}(M, Z, E_\nu)}{{\rm d}E_{\nu}} \nonumber \\
 & & = \bar{P}\,\frac{{\rm d}N_{\bar{\nu}_e}^{0}(M, Z, E_\nu)}{{\rm d}E_{\nu}} + (1-\bar{P})\,\frac{{\rm d}N_{\bar{\nu}_x}^{0}(M, Z, E_\nu)}{{\rm d}E_{\nu}},
\label{eq:srvprob}
\end{eqnarray}
where $\bar{P}_{\alpha e}$ ($\alpha = e, \mu, \tau$) is a conversion probability from $\bar\nu_\alpha$ to $\bar\nu_e$ satisfying $\bar{P}_{ee} + \bar{P}_{\mu e} + \bar{P}_{\tau e}=1$, and $\bar{P}\,(=\bar{P}_{ee})$ is the survival probability of $\bar{\nu}_e$ passing through stellar envelopes and space \citep[][]{dighe00}. Meanwhile, ${\rm d}N_{\bar{\nu}_\alpha}^{0}(M, Z, E_\nu)/{\rm d}E_\nu$ is the original spectrum of $\bar{\nu}_\alpha$. Note that the original spectra of $\bar{\nu}_{\mu}$ and $\bar{\nu}_{\tau}$ are almost identical because $\bar{\nu}_{\mu}$ and $\bar{\nu}_{\tau}$ do not have charged-current reactions around the neutrino-emitting surface of a supernova core and, therefore, both of them can be denoted as ${\rm d}N_{\bar{\nu}_x}^{0}(M, Z, E_\nu)/{\rm d}E_\nu$. These original spectra are taken from the Supernova Neutrino Database \citep[][]{self13a}.

The flavor eigenstates of antineutrinos, $|\bar{\nu}_{\alpha}\rangle$, are linear combinations of the energy eigenstates in vacuum, $|\bar{\nu}_{i}\rangle$ ($i=1,2,3$ with neutrino masses $m_1, m_2, m_3$, respectively):
\begin{equation}
|\bar{\nu}_{\alpha}\rangle = \sum_i U_{\alpha i}^* |\bar{\nu}_{i}\rangle,
\label{eq:neuvec}
\end{equation}
where the mixing matrix $U$ can be expressed as
\begin{eqnarray}
& &  U_{\alpha i} = \nonumber \\
& &  \left(
  \begin{array}{@{\,}ccc@{\,}}
    c_{12} c_{13}                                       & s_{12} c_{13}                                      & s_{13} e^{-i \delta} \\
    - s_{12} c_{23} - c_{12} s_{23} s_{13} e^{i \delta} & c_{12} c_{23} - s_{12} s_{23} s_{13} e^{i \delta}  & s_{23} c_{13}        \\
    s_{12} s_{23} - c_{12} c_{23} s_{13} e^{i \delta}   & -c_{12} s_{23} - s_{12} c_{23} s_{13} e^{i \delta} & c_{23} c_{13}
  \end{array}
  \right), \nonumber \\
& &
\label{eq:mixmtrx}
\end{eqnarray}
with mixing angles $\theta_{12},\theta_{23},\theta_{13}$, CP phase $\delta$ (we set $\delta=0$ in this study) and $s_{ij}=\sin\theta_{ij}, c_{ij}=\cos\theta_{ij}$. The three mixing angles have been measured in several experiments/observations: $\sin^{2}\theta_{12}\sim0.31$, $\sin^{2} \theta_{23}\sim0.44$, $\sin^{2}\theta_{13} \sim 0.02$ \citep[][]{pdg14}. Using the matrix elements of $U$, the number spectrum of $\bar{\nu}_e$ is denoted as \citep[][]{dighe00}
\begin{eqnarray}
\frac{{\rm d}N_{\bar{\nu}_e}}{{\rm d}E_{\nu}}&=&|U_{e1}|^2\frac{{\rm d}N_{\bar{\nu}_1}}{{\rm d}E_{\nu}}
                     +|U_{e2}|^2\frac{{\rm d}N_{\bar{\nu}_2}}{{\rm d}E_{\nu}}
                     +|U_{e3}|^2\frac{{\rm d}N_{\bar{\nu}_3}}{{\rm d}E_{\nu}} \nonumber \\ 
&=& \cos^{2}\theta_{12}\cos^{2}\theta_{13}\frac{{\rm d}N_{\bar{\nu}_1}}{{\rm d}E_{\nu}}
 +  \sin^{2}\theta_{12}\cos^{2}\theta_{13}\frac{{\rm d}N_{\bar{\nu}_2}}{{\rm d}E_{\nu}} \nonumber \\
& & +  \sin^{2}\theta_{13}\frac{{\rm d}N_{\bar{\nu}_3}}{{\rm d}E_{\nu}},
\label{eq:neuosc}
\end{eqnarray}
where ${\rm d}N_{\bar{\nu}_i}/{\rm d}E_{\nu}$ is the number spectrum of $\bar{\nu}_{i}$.

In the supernova core, a good approximation is that the lowest neutrino energy eigenstate corresponds to $\bar{\nu}_e$ and other states are mixtures of $\bar{\nu}_{\mu}$ and $\bar{\nu}_{\tau}$. Furthermore, if the stellar density profiles are not sufficiently steep to break up the adiabaticity of propagating eigenstates, neutrinos emitted from the dense core pass through the stellar envelope with varying density along a single energy eigenstate into vacuum space. Note that there is still uncertainty concerning the mass hierarchy: normal ($m_1 < m_2 < m_3$) or inverted ($m_3 < m_1 < m_2$). Thus, we obtain ${\rm d}N_{\bar{\nu}_1}/{\rm d}E_{\nu}\sim{\rm d}N_{\bar{\nu}_e}^{0}/{\rm d}E_{\nu}$, ${\rm d}N_{\bar{\nu}_2}/{\rm d}E_{\nu}\sim{\rm d}N_{\bar{\nu}_x}^{0}/{\rm d}E_{\nu}$, ${\rm d}N_{\bar{\nu}_3}/{\rm d}E_{\nu}\sim{\rm d}N_{\bar{\nu}_x}^{0}/{\rm d}E_{\nu}$ for the case of the normal mass hierarchy and ${\rm d}N_{\bar{\nu}_3}/{\rm d}E_{\nu}\sim{\rm d}N_{\bar{\nu}_e}^{0}/{\rm d}E_{\nu}$, ${\rm d}N_{\bar{\nu}_1}/{\rm d}E_{\nu}\sim{\rm d}N_{\bar{\nu}_x}^{0}/{\rm d}E_{\nu}$, ${\rm d}N_{\bar{\nu}_2}/{\rm d}E_{\nu}\sim{\rm d}N_{\bar{\nu}_x}^{0}/{\rm d}E_{\nu}$ for the case of the inverted mass hierarchy \citep[][]{dighe00}. Substituting these relations into Equation~(\ref{eq:neuosc}), we obtain
\begin{equation}
\frac{{\rm d}N_{\bar{\nu}_e}}{{\rm d}E_{\nu}} \sim 0.68\,\frac{{\rm d}N_{\bar{\nu}_e}^{0}}{{\rm d}E_{\nu}} + 0.32\,\frac{{\rm d}N_{\bar{\nu}_x}^{0}}{{\rm d}E_{\nu}},
\label{eq:neuoscnh}
\end{equation}
and $\bar{P}=0.68$ for the normal mass hierarchy. For the inverted mass hierarchy, Equation~(\ref{eq:neuosc}) becomes
\begin{equation}
\frac{{\rm d}N_{\bar{\nu}_e}}{{\rm d}E_{\nu}} \sim \frac{{\rm d}N_{\bar{\nu}_x}^{0}}{{\rm d}E_{\nu}},
\label{eq:neuoscih}
\end{equation}
where we neglect the small value of $\sin^2\theta_{13}\sim0.02$. This means a complete transition ($\bar{P}=0$). Incidentally, we do not take into account the breaking of adiabaticity due to the shock wave propagation \citep[e.g.,][]{gala10}, which should be minor for a time-integrated signal \citep[][]{kawa10}. We also neglect neutrino-neutrino collective effects \citep[e.g.,][]{chakr11,dasgu12}, whose contribution to SRNs is estimated to be about 5-10\% \citep[][]{luna12}.

\section{Models of Galaxy Evolution}\label{sec:gal}

To derive a realistic spectrum of SRNs, it is mandatory to use a reliable core-collapse rate based on models of galaxy evolution. Here we consider the stellar mass, $M_\ast$, SFR, $\dot M_\ast$, and metallicity, 12+log$_{10}$(O/H), of galaxies. The SFR of galaxies is responsible for the CSFRD, $\dot \rho_\ast (z)$, and the metallicity of galaxies corresponds to that of the progenitors in them. Furthermore, $M_\ast$, $\dot M_\ast$ and 12+log$_{10}$(O/H) are related to each other \citep[e.g.,][]{brinch04,trem04,salim07,mannu10,mannu11,niino12}. In this study, for our reference model, we adopt the galaxy stellar mass function and mass-dependent SFR proposed by \citet{da08} (hereafter DA08) as functions of redshift, and the redshift-dependent mass-metallicity relation from M08 is utilized. We also consider other models for comparison.

\subsection{Cosmic Star Formation Rate Density}\label{sec:csfrd}

\begin{figure}
\plotone{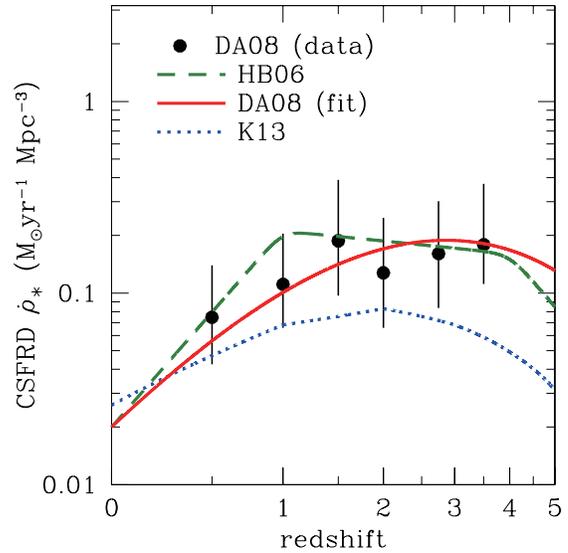}
\caption{CSFRD as a function of redshift. Dashed, solid and dotted lines correspond to the models in HB06, DA08 and K13, respectively. Plots are calculated from the data in Tables~1 and 2 in DA08.}
\label{fig:csfrd}
\end{figure}

In DA08, the redshift evolution of the stellar mass function is presented on the basis of the data spanning $0<z<5$ \citep[][]{drory05}. It is assumed to have a Schechter form and is written as
\begin{eqnarray}
\phi_{\rm SMF}(M_\ast,z) \ {\rm d}M_\ast & = & \phi_0(z) \left(\frac{M_\ast}{M_0^{\rm DA08}(z)}\right)^{\alpha^{\rm DA08}} \times \nonumber \\
 & & \exp\left(-\frac{M_\ast}{M_0^{\rm DA08}(z)}\right)\frac{{\rm d}M_\ast}{M_0^{\rm DA08}(z)}, \nonumber \\
 & &
\label{eq:smf}
\end{eqnarray}
where the best-fitting parameterization is
\begin{subequations}
\begin{equation}
\phi_0(z) = 0.0031 \times (1+z)^{-1.07} \ {\rm Mpc}^{-3} \ {\rm dex}^{-1},
\label{eq:smf-phi0}
\end{equation}
\begin{equation}
\log_{10} \left( \frac{M_0^{\rm DA08}(z)}{M_\odot} \right) = 11.35 - 0.22 \times \ln (1+z),
\label{eq:smf-m0}
\end{equation}
\begin{equation}
\alpha^{\rm DA08} = -1.3.
\label{eq:smf-alpha}
\end{equation}
\label{eq:smf-spl}
\end{subequations}
The redshift dependence of SFR is also parameterized in DA08 as a function of stellar mass:
\begin{eqnarray}
\dot M_\ast(M_\ast,z) & = & \dot M^0_\ast(z) \left(\frac{M_\ast}{M_1^{\rm DA08}(z)}\right)^{\beta^{\rm DA08}} \times \nonumber \\
 & & \exp\left(-\frac{M_\ast}{M_1^{\rm DA08}(z)}\right),
\label{eq:sfr}
\end{eqnarray}
with
\begin{subequations}
\begin{equation}
M_1^{\rm DA08}(z) = 2.7 \times 10^{10} \times (1+z)^{2.1} \ M_\odot,
\label{eq:sfr-m1}
\end{equation}
\begin{equation}
\beta^{\rm DA08} = 0.6,
\label{eq:sfr-beta}
\end{equation}
\label{eq:sfr-spl}
\end{subequations}
while the fitting form of $\dot M^0_\ast(z)$ is not given in DA08. In our study, we adopt the following functional form:
\begin{equation}
\dot M^0_\ast(z) = \dot M^{\rm N15}_{\ast,0} (1+z)^{\alpha^{\rm N15}} \exp(\beta^{\rm N15}z) ,
\label{eq:sfr-mdot0}
\end{equation}
with $\dot M^{\rm N15}_{\ast,0}=1.183M_\odot \ {\rm yr}^{-1}$, $\alpha^{\rm N15}=5.5$ and $\beta^{\rm N15} = -0.78$, so as to fit the CSFRD, $\dot \rho_\ast (z)$, of the data in DA08 shown in Figure~\ref{fig:csfrd}. Note that CSFRD is obtained from
\begin{equation}
\dot \rho_\ast(z) = \int^\infty_0 \dot M_\ast(M_\ast,z) \phi_{\rm SMF}(M_\ast,z) \ {\rm d}M_\ast ,
\label{eq:csfrd}
\end{equation}
and the value of $\dot M^{\rm N15}_{\ast,0}$ is determined to obtain $\dot \rho_\ast(0) = 0.02 M_\odot~{\rm yr}^{-1}~{\rm Mpc}^{-3}$, which is consistent with the values in recent studies of $0.017 M_\odot~{\rm yr}^{-1}~{\rm Mpc}^{-3}$ \citep[][]{hori11}, $0.026 M_\odot~{\rm yr}^{-1}~{\rm Mpc}^{-3}$ (K13) and 0.007-$0.02 M_\odot~{\rm yr}^{-1}~{\rm Mpc}^{-3}$ \citep[][]{mathe14}. 

So far, the CSFRD proposed by \citet{hb06} (hereafter HB06) has often been used for the calculation of the SRN spectrum. By performing a modification using data of gamma ray bursts for $z>4$, \citet{yuk08} obtained the parametric fit
\begin{eqnarray}
\dot{\rho}^{\rm HB06}_\ast (z) & = & \dot{\rho}_0 \Biggl[ (1+z)^{\alpha^{\rm HB06} \eta^{\rm HB06}} + \left( \frac{1+z}{B} \right)^{\beta^{\rm HB06} \eta^{\rm HB06}} \nonumber \\
 & &  + \left( \frac{1+z}{C} \right)^{\gamma^{\rm HB06} \eta^{\rm HB06}} \Biggr]^{1/\eta^{\rm HB06}},
\label{eq:yuksfr}
\end{eqnarray}
with $\alpha^{\rm HB06} = 3.4$, $\beta^{\rm HB06} = -0.3$, $\gamma^{\rm HB06} = -3.5$, $\eta^{\rm HB06} = -10$ and $\dot{\rho}_0 = 0.02 M_\odot~{\rm yr}^{-1}~{\rm Mpc}^{-3}$. In this form, breaks are made at $z_1=1$ and $z_2=4$ by coefficients $B=(1+z_1)^{1-\alpha^{\rm HB06}/\beta^{\rm HB06}}$ and $C=(1+z_1)^{(\beta^{\rm HB06}-\alpha^{\rm HB06})/\gamma^{\rm HB06}}(1+z_2)^{1-\beta^{\rm HB06}/\gamma^{\rm HB06}}$, respectively. While various observational results were compiled in HB06, there are uncertainties in the dust obscuration correction and the conversion from UV luminosity to SFR. In fact, most of the theoretical models for galaxy formation underpredict CSFRD at $z \gtrsim 1$ and, according to K13, this inconsistency is due to these observational uncertainties. Therefore, for the SRN spectrum, we also examine CSFRD based on the semi-analytic model of galaxy formation presented in K13 as well as the observational model in HB06. The CSFRD models considered in this paper are compared in Figure~\ref{fig:csfrd}. We can see that our reference model of CSFRD based on DA08 lies between those in HB06 and K13 for $z \lesssim 2$, the region mainly responsible for SRNs.

With the CSFRD models stated above, we obtain the total core-collapse rate, $R_{\rm CC} (z)$, as
\begin{equation}
R_{\rm CC}(z) = \dot \rho_\ast(z) \times \frac{\int^{M_{\rm max}}_{M_{\rm min}} \psi_{\rm IMF} (M) \ {\rm d}M}{\int^{100M_\odot}_{0.1M_\odot} M\psi_{\rm IMF} (M) \ {\rm d}M},
\label{eq:ccrate}
\end{equation}
adopting the Salpeter initial mass function ($\psi_{\rm IMF} (M) \propto M^{-2.35}$) with a mass range of 0.1-$100M_\odot$ as in DA08. Here $M_{\rm max}$ and $M_{\rm min}$ are the maximum and minimum masses of progenitors that end with a core collapse, respectively. Their values are given later in \S~\ref{sec:snn}.

\subsection{Metallicity Evolution}\label{sec:metal}

\begin{figure*}
\plotone{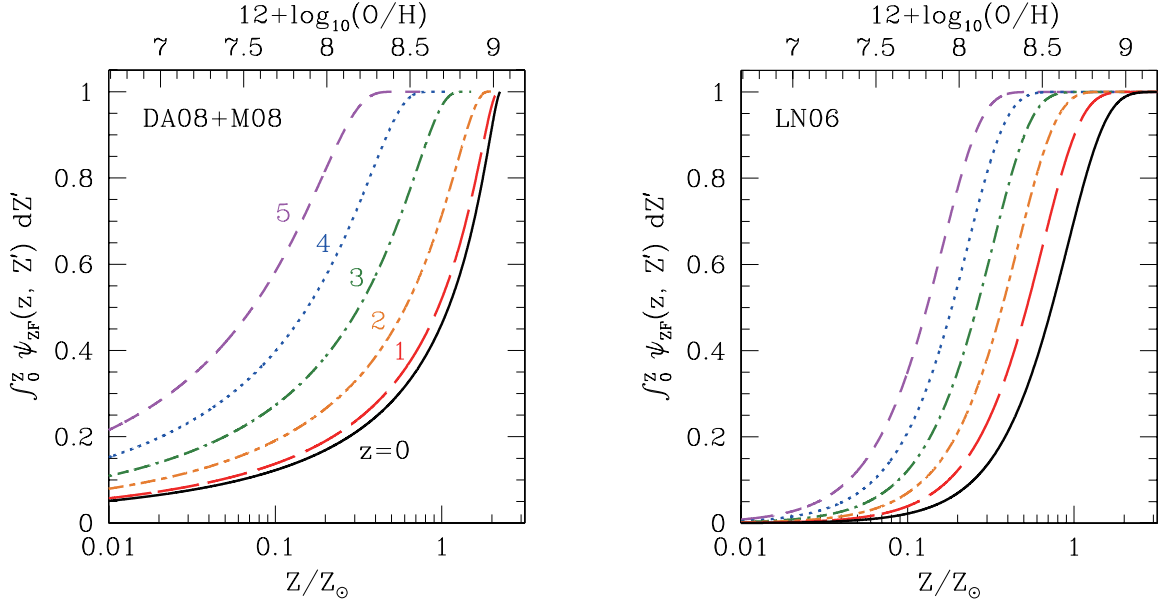}
\caption{Normalized cumulative metallicity distribution function, which represents the fraction of progenitors with metallicity less than $Z$, for the models in DA08$+$M08 ({\it left}) and LN06 ({\it right}). The lines correspond, from bottom to top, to redshifts of $z=0$, 1, 2, 3, 4 and 5.}
\label{fig:metafun}
\end{figure*}

The redshift dependence of the galaxy mass-metallicity relation was investigated in M08. The following analytical form was adopted:
\begin{eqnarray}
12+\log_{10}({\rm O}/{\rm H}) & = & -0.0864 \times (\log_{10}M_\ast -\log_{10}M_0^{\rm M08})^2  \nonumber \\
 & & + K_0^{\rm M08} ,
\label{eq:M08}
\end{eqnarray}
where the best fit values of $\log_{10}M_0^{\rm M08}$ and $K_0^{\rm M08}$ at different redshifts are listed in Table~5 of M08. While there are two choices for the values at $z=3.5$ with different models of spectral synthesis, we adopt case {\it a} based on \citet{bc03}. Using linear interpolation and extrapolation, we obtain the mass-metallicity relation for redshifts ranging from 0 to 5. We assume that the solar metallicity ($Z_\odot=0.02$) corresponds to the oxygen abundance of $12+\log_{10}({\rm O}/{\rm H}) = 8.69$ \citep[][]{alle01}, i.e.,
\begin{equation}
\log_{10}\left(\frac{Z}{Z_\odot}\right) = 12+\log_{10}({\rm O}/{\rm H}) -8.69.
\label{eq:solmet}
\end{equation}
In our study, we assume that the stellar metallicity is identical to the value for its host galaxy for simplicity, whereas the metallicity variation inside a galaxy is currently being discussed \citep[e.g.,][]{leve11,niino11,sanders12,taddia13,niino14}. We combine the mass-metallicity relation with the stellar mass function (\ref{eq:smf}) and SFR (\ref{eq:sfr}) of DA08 to derive the metallicity distribution function of progenitors $\psi_{\rm ZF} (z, Z)$ as
\begin{eqnarray}
 & & \int^{Z}_0 \psi_{\rm ZF}(z,Z^\prime) \ {\rm d}Z^\prime  \nonumber \\ 
 & & = \frac{\displaystyle \int^{M_\ast(z,Z)}_0 \dot M_\ast(M_\ast^\prime,z) \phi_{\rm SMF}(M_\ast^\prime,z) \ {\rm d}M_\ast^\prime}{\displaystyle \int^\infty_0 \dot M_\ast(M_\ast^\prime,z) \phi_{\rm SMF}(M_\ast^\prime,z) \ {\rm d}M_\ast^\prime},
\label{eq:psizf}
\end{eqnarray}
where $M_\ast(z,Z)$ is the stellar mass of a galaxy with metallicity $Z$ at redshift $z$ calculated from Equations~(\ref{eq:M08}) and (\ref{eq:solmet}). The left-hand side of Equation~(\ref{eq:psizf}), which is shown in Figure~\ref{fig:metafun}, represents the fraction of progenitors with metallicity less than $Z$. Hereafter, we refer to this model as DA08+M08.

In this study, we also consider the model of metallicity evolution proposed by \citet{ln06} (hereafter LN06). According to their model, the metallicity distribution function is given as
\begin{eqnarray}
\int^{Z}_0 \psi_{\rm ZF}^{\rm LN06}(z,Z^\prime) \ {\rm d}Z^\prime & = & \frac{\displaystyle \int^{M_\ast(z,Z)}_0 M_\ast^\prime \phi_{\rm SMF}^{\rm LN06}(M_\ast^\prime) \ {\rm d}M_\ast^\prime}{\displaystyle \int^\infty_0 M_\ast^\prime \phi_{\rm SMF}^{\rm LN06}(M_\ast^\prime) \ {\rm d}M_\ast^\prime} \nonumber \\
& = & \frac{\hat \Gamma(\alpha^{\rm LN06}+2, M_\ast(z,Z)/M_0^{\rm LN06})}{\Gamma(\alpha^{\rm LN06}+2)}, \nonumber \\
& &
\label{eq:LN06}
\end{eqnarray}
where $\hat \Gamma$ and $\Gamma$ are the incomplete and complete gamma functions originating from the Schechter form of their stellar mass function $\phi_{\rm SMF}^{\rm LN06}(M_\ast) \propto (M_\ast/M_0^{\rm LN06})^{\alpha^{\rm LN06}}\exp(-M_\ast/M_0^{\rm LN06})$, respectively. Note that, in Equation (\ref{eq:LN06}), $M_\ast \phi_{\rm SMF}$ is integrated instead of $\dot M_\ast \phi_{\rm SMF}$ in Equation (\ref{eq:psizf}). However, $\dot M_\ast \phi_{\rm SMF}$ is reasonable for our purpose because short-lived massive progenitors contribute to SRNs. Furthermore, in LN06, the mass-metallicity relation was assumed to be
\begin{equation}
\frac{M_\ast(z,Z)}{M_0^{\rm LN06}} = K^{\rm LN06}\left(\frac{Z}{Z_\odot}\right)^{\beta^{\rm LN06}}10^{\gamma^{\rm LN06}\beta^{\rm LN06}z},
\label{eq:LN06MZ}
\end{equation}
where $\gamma^{\rm LN06}$ is related to the average cosmic metallicity scaling as 
\begin{equation}
\frac{{\rm d}}{{\rm d}z} \log_{10}\left(\frac{Z}{Z_\odot}\right) = -\gamma^{\rm LN06}.
\label{eq:LN06Zsca}
\end{equation}
Taking the numbers in LN06, $\alpha^{\rm LN06}=-1.16$, $\beta^{\rm LN06}=2$, $\gamma^{\rm LN06}=0.15$ and $K^{\rm LN06}=1$, we obtain the metallicity distribution function (Figure~\ref{fig:metafun}).

\section{Neutrino Emission from Stellar Core Collapse}\label{sec:snn}

\begin{figure*}
\plotone{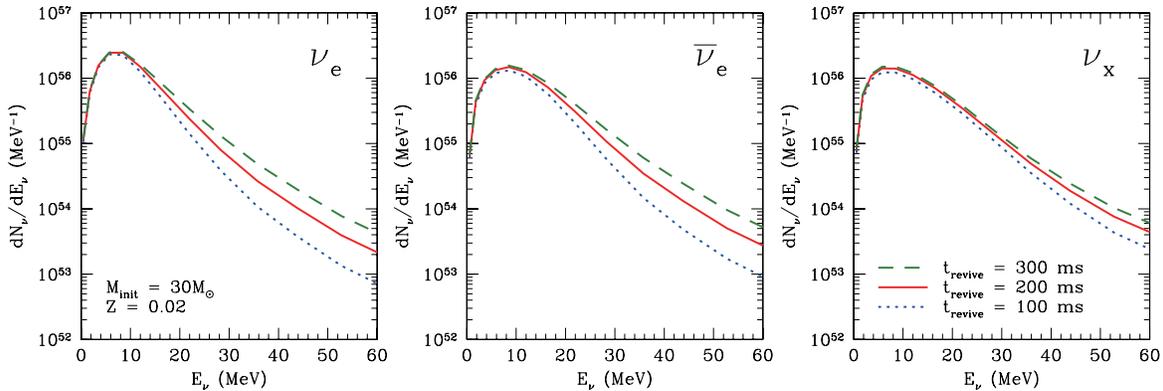}
\caption{Neutrino number spectra of supernova with $30M_\odot$, $Z = 0.02$ and shock revival times of $t_{\rm revive}=100$~ms (dotted), 200~ms (solid) and 300~ms (dashed). The left, central and right panels correspond to $\nu_e$, $\bar\nu_e$ and $\nu_x$ ($=\nu_\mu=\bar \nu_\mu=\nu_\tau=\bar \nu_\tau$), respectively.}
\label{fig:snspc}
\end{figure*}

For the calculation of the SRN spectrum, we utilize the numerical results for the neutrino spectra emitted from various progenitors in the Supernova Neutrino Database \citep[][]{self13a}. The progenitor models in this data set are computed by a Henyey-type stellar evolution code \citep[][]{umeda12} and models with the initial masses $M= 13M_\odot$, $20M_\odot$, $30M_\odot$ and $50M_\odot$ are provided. In our model, we set $M_{\rm min}=10M_\odot$ in Equation~(\ref{eq:ccrate}) for the minimum mass of progenitors that end with a core collapse for consistency with the evolutionary calculation of the progenitors. We do not take into account the contribution due to electron-capture supernovae of low-mass progenitors \citep[][]{luna12,mathe14}. Note that recent theoretical studies suggest that their mass range is narrow \citep[e.g.,][]{poe08,dohe15}. On the other hand, the maximum mass of progenitors is $M_{\rm max}=100M_\odot$ from the mass range of the initial mass function. Furthermore, in the Supernova Neutrino Database, two different values of stellar metallicity ($Z=0.02$ and 0.004, i.e., the solar abundance and one-fifth of the solar abundance, respectively) are considered, and we consider the metallicity distribution for SRNs using a threshold of $Z=\sqrt{0.02 \times 0.004}$. Therefore, eight progenitor models are adopted in total. As shown in Equation~(\ref{eq:srn}), their neutrino spectra, ${\rm d}N(M, Z, E^\prime_\nu)/{\rm d}E^\prime_\nu$, are integrated in SRNs.

\subsection{Supernova Explosion and Shock Revival Time}\label{sec:sn}

The physical factors that cause the CCSN explosion are not well understood \citep[e.g.,][]{janka12,kotake12,burrows13}. However, it is widely accepted that the shock wave launched by the bounce due to the nuclear repulsion stalls once but is revived by some mechanism. The process leading to shock revival is still a matter of debate. In the Supernova Neutrino Database, the shock revival time $t_{\rm revive}$ is introduced as a parameter that reflects the unknown explosion mechanism. In this data set, under spherical symmetry, neutrino-radiation hydrodynamic simulations \citep[][]{yamada97,yamada99,sumi05} and quasi-static evolutionary calculations of neutrino diffusion \citep[][]{suzuki94} are used for the early and late phases of the supernova explosion, respectively. Although the neutrino-radiation hydrodynamic simulations do not lead to a natural supernova explosion, they are phenomenologically connected to quasi-static evolutionary calculations of neutrino diffusion, based on physical considerations \citep[see Figure~14 of][]{self13a}. The shock revival time after the bounce has been estimated to be on the order of 100~ms. For instance, according to \citet{belcz12}, it is preferably as short as 100-200~ms to account for the observed mass distributions of neutron stars and black holes. On the other hand, \citet{yama13} suggested from a numerical study that the shock relaunch should be delayed until 300-400~ms to simultaneously produce the appropriate explosion energy and nickel yields. Therefore, in our study, the shock wave is assumed to be revived at either $t_{\rm revive}=100$, 200 or 300~ms after the bounce.

In Figure~\ref{fig:snspc}, we show the neutrino number spectra of supernovae with different shock revival times for the models with $30M_\odot$ and $Z = 0.02$ in the Supernova Neutrino Database. The total emission number and energy of supernova neutrinos depend on the shock revival time; they increase with $t_{\rm revive}$ because more material accretes onto the collapsed core, releasing a huge amount of gravitational potential energy. In particular, the difference is larger in the high-energy regime. This is because high-energy neutrinos are mainly emitted in the early phase when the proto-neutron star continues to be heated as a result of mass accretion. After the shock revival, the accretion stops and the mean energy of emitted neutrinos decreases, which is called the cooling phase. To estimate the uncertainty of the explosion mechanism, we calculate the spectrum of SRNs for models with different values of $t_{\rm revive}$ \citep[][]{self13c}. Note that the shock revival time may not be the same for all progenitors but be longer for more massive progenitors. We may underestimate the SRN flux using models with $t_{\rm revive}=100$~ms especially for massive progenitors. Therefore, in this study, we regard the cases with $t_{\rm revive}=300$ and 100~ms as an indication of upper and lower bounds, respectively.

\subsection{Black Hole Formation and Nuclear Equation of State}\label{sec:bh}

\begin{figure}[b]
\plotone{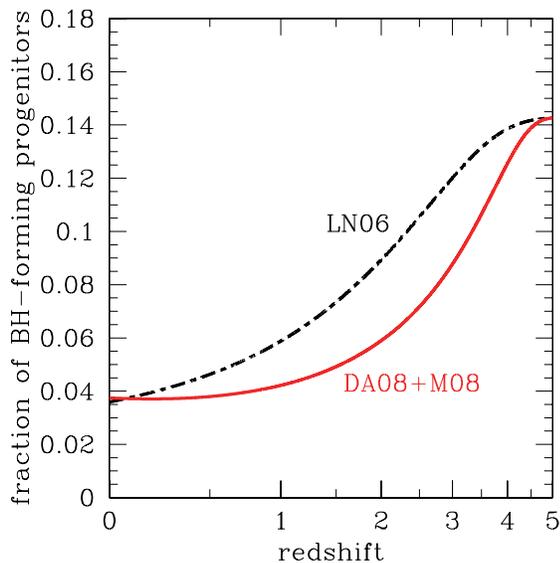}
\caption{Fraction of black-hole-forming progenitors as a function of redshift. Dot-dashed and solid lines correspond to the models with the metallicity evolution of LN06 and DA08$+$M08, respectively.}
\label{fig:frabh}
\end{figure}

Some massive stars are thought to fail to produce supernovae, leaving a black hole as a remnant \citep[e.g.,][]{lieb04,sumi06,ocon10}. The failed supernovae also emit neutrinos from the bounce to black hole formation and, therefore, contribute to the overall flux of SRNs \citep[][]{luna09,lien10,self13c}. In the Supernova Neutrino Database, the progenitor model with $30M_\odot$ and $Z = 0.004$ is assumed to become a failed supernova because of its high core mass. Note that the models with $Z=0.004$ have higher core mass than those with $Z=0.02$ because the mass loss rate is larger for higher metallicity. On the other hand, the core mass is not monotonically related to the initial mass of progenitors because the mass loss rate is larger for a higher mass. Thus, the core mass of the model with the initial mass of $30M_\odot$ is the highest for the progenitors in the Supernova Neutrino Database. In contrast, progenitor models other than $(M_\mathrm{init}, Z)=(30M_\odot, 0.004)$ are assumed to be ordinary supernovae. Combining this assumption with the Salpeter initial mass function $\psi_{\rm IMF} (M)$ and the metallicity distribution function $\psi_{\rm ZF} (z, Z)$ of Equation~(\ref{eq:psizf}) or (\ref{eq:LN06}), we obtain the fraction of black-hole-forming progenitors and calculate the spectrum of SRNs including failed supernovae. In Figure~\ref{fig:frabh}, the fraction of black-hole-forming progenitors is shown as a function of redshift for our models. At present, its value is hardly constrained by the observational data, while \citet{hori14} estimated it to be 0.2-0.4 on the basis of another progenitor set \citep[][]{woosley02}.

\begin{deluxetable*}{lcccccccc}
\tabletypesize{\footnotesize}
\tablewidth{0pt}
\tablecaption{Numerical results for black hole formation of progenitor with $(M, Z) = (30M_\odot, 0.004)$.}
\tablehead{ & $t_\mathrm{BH}$ & $\langle{E_{\nu_e}\rangle}$ & $\langle{E_{\bar \nu_e}\rangle}$ & $\langle{E_{\nu_x}\rangle}$ & $E_{\nu_e,\mathrm{tot}}$ & $E_{\bar \nu_e,\mathrm{tot}}$ & $E_{\nu_x,\mathrm{tot}}$ & $E_{\nu_\mathrm{all},\mathrm{tot}}$ \\
 EOS & (ms) & (MeV) & (MeV) & (MeV) & ($10^{52}$~erg) & ($10^{52}$~erg) & ($10^{52}$~erg) & ($10^{53}$~erg) }
\startdata
 Shen        & 842 & 17.5 & 21.7 & 23.4 & 9.49 & 8.10 & 4.00 & 3.36 \\
 LS(220~MeV) & 342 & 12.5 & 16.4 & 22.3 & 4.03 & 2.87 & 2.11 & 1.53
\enddata
\label{tab:bh}
\tablecomments{$t_\mathrm{BH}$ is the time to black hole formation measured from the core bounce. The mean energy of the emitted $\nu_i$ until black hole formation is denoted as $\langle{E_{\nu_i}\rangle} \equiv E_{\nu_i,\mathrm{tot}}/N_{\nu_i,\mathrm{tot}}$, where $E_{\nu_i,\mathrm{tot}}$ and $N_{\nu_i,\mathrm{tot}}$ are the total energy and number of neutrinos, respectively. $\nu_x$ stands for $\mu$- and $\tau$-neutrinos and their anti-particles: $E_{\nu_x}=E_{\nu_\mu}=E_{\bar{\nu}_\mu}=E_{\nu_\tau}=E_{\bar{\nu}_\tau}$. $E_{\nu_\mathrm{all},\mathrm{tot}}$ is the total neutrino energy summed over all species.}
\end{deluxetable*}

\begin{figure*}
\plotone{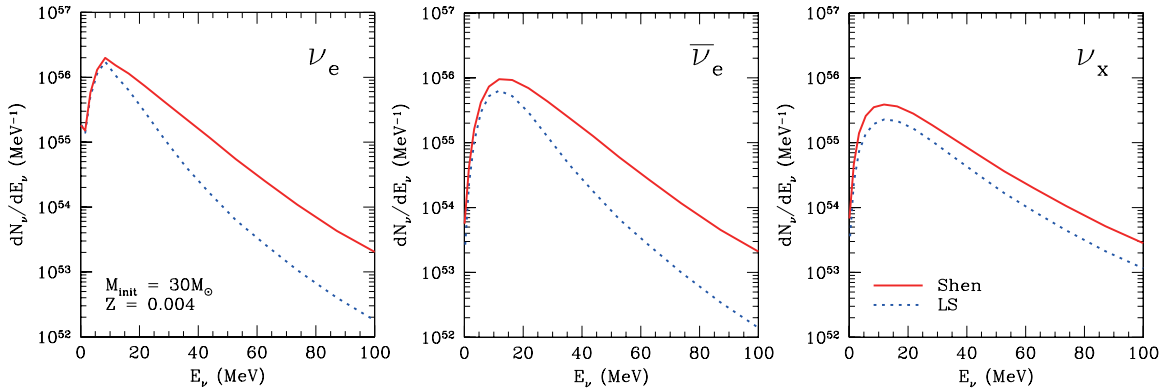}
\caption{Neutrino number spectra for black hole formation with $30M_\odot$, $Z = 0.004$ and Shen EOS (solid) and LS EOS (dotted). The left, central and right panels correspond to $\nu_e$, $\bar\nu_e$ and $\nu_x$ ($=\nu_\mu=\bar \nu_\mu=\nu_\tau=\bar \nu_\tau$), respectively.}
\label{fig:bhspc}
\end{figure*}

Although the neutrino signal from a failed supernova is sensitive to the nuclear EOS \citep[][]{sumi06}, there is a result based on a single model by \citet{Shen:1998gq,Shen:1998by} in the Supernova Neutrino Database. To estimate the uncertainty of EOS, in this study, we compute the neutrino signal from the progenitor model with $30M_\odot$ and $Z = 0.004$ using another EOS by \citet{Lattimer:1991nc} (hereafter LS). While the LS EOS has three choices with different values of incompressibility $K$, we adopt the set with $K=220$~MeV. The method of computation is fully consistent with the original one in the Supernova Neutrino Database \citep[][]{self13a}. Incidentally, the EOS uncertainty of neutrino flux from ordinary CCSN has been evaluated for both the accretion phase \citep[][]{sumi05} and the cooling phase \citep[][]{suzuki05} to be $\sim$10\%, which is smaller than the dispersion due to the shock revival time in the Supernova Neutrino Database ($\sim$20-50\%).

We follow the neutrino emission until black hole formation with the neutrino-radiation hydrodynamic simulation. Note that the shock wave is not revived in this model and the mass accretion continues until black hole formation. Therefore, a larger amount of matter is accreted and more neutrinos are emitted compared with an ordinary supernova. Furthermore, the mean energy of neutrinos emitted from a failed supernova is higher than that from an ordinary supernova because the heating due to the accretion continues. It should also be noted that the neutrino emission from a failed supernova is rich in $\nu_e$ and $\bar\nu_e$. This is because $\nu_e$ and $\bar\nu_e$ are emitted more abundantly than $\nu_x$ ($=\nu_\mu=\bar \nu_\mu=\nu_\tau=\bar \nu_\tau$) from the accretion of matter owing to the capture of electrons and positrons on nucleons. In contrast, during the cooling of a proto-neutron star, neutrinos of all species are emitted equivalently.

In Figure~\ref{fig:bhspc} and Table~\ref{tab:bh}, the results for the Shen EOS and LS EOS are compared. We find that the model with the Shen EOS produces more neutrinos than that with the LS EOS. This is for the following reason. The Shen EOS with the incompressibility of $K=281$~MeV is stiffer than the LS EOS adopted in this study. Thus, the maximum mass of neutron stars is higher and the amount of mass accretion is larger for the Shen EOS. In fact, the model with the Shen EOS takes a longer time to form a black hole (Table~\ref{tab:bh}). As a result, more potential energy can be converted into the emission energy of neutrinos. These features are consistent with previous studies \citep[][]{sumi06,fischer09,self10a}.

\section{Expected Event Rate at Super-Kamiokande}\label{sec:event}

In this section, we present the results for the spectrum of SRNs and evaluate the event rate for Super-Kamiokande over 1 year. For this, we only consider the inverse $\beta$ decay reaction of electron antineutrinos,
\begin{equation}
\bar{\nu}_e + p \to e^+ + n,
\label{eq:ibd}
\end{equation}
which is the most promising channel for detection. For Super-Kamiokande with a 22.5~kton fiducial volume, the number of target protons is set to $N_t=1.5 \times 10^{33}$. Note that, in the following, we do not take into account the detection efficiency for simplicity. Thus, using different normalization, our results are applicable to different detectors with the inverse $\beta$ decay events. The event rate spectrum for positrons is written as
\begin{equation}
\frac{{\rm d}N_{e^+}(E_{e^+})}{{\rm d}E_{e^+}} = N_t \sigma(E_{\bar{\nu}_e}) \frac{{\rm d}F(E_{\bar{\nu}_e})}{{\rm d}E_{\bar{\nu}_e}},
\label{eq:convcoef}
\end{equation}
where $\sigma(E_{\bar{\nu}_e})$ is the cross section for the inverse $\beta$ decay using the approximation shown in Equation~(25) of \citet{stru03}. The positron energy is $E_{e^+}=E_{\bar{\nu}_e}-\Delta c^2$ with neutron-proton mass difference $\Delta$. In the following, we choose a model with the CSFRD of DA08, the metallicity distribution function of DA08+M08, a shock revival time of $t_{\rm revive}=200$~ms and the Shen EOS as a reference model.

~\\

\subsection{Impact of Metallicity Evolution and Black Hole Formation}\label{sec:mebh}

\begin{figure*}
\plotone{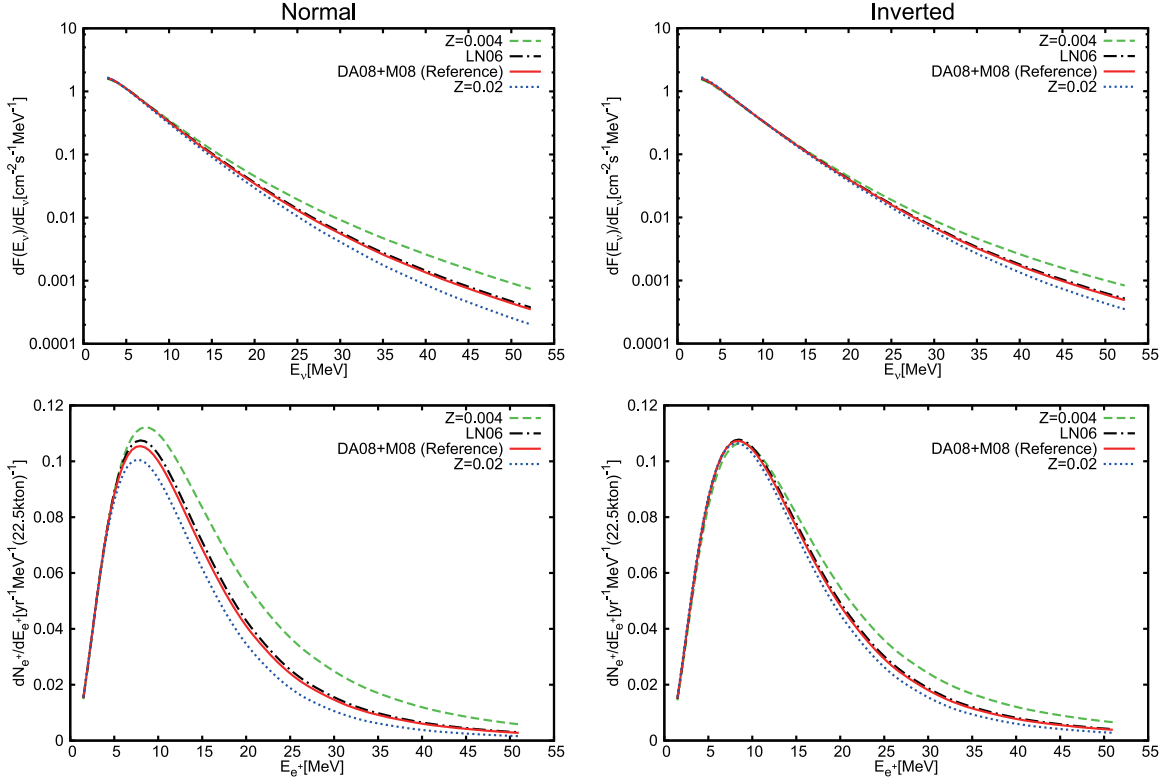}
\caption{Fluxes of SRNs ({\it upper panels}) and event rate spectra in Super-Kamiokande over 1 year ({\it lower panels}) obtained using models with CSFRD of DA08, shock revival time of $t_{\rm revive}=200$~ms and Shen EOS. The left and right panels show the results for the normal and inverted mass hierarchies, respectively. Solid and dot-dashed lines correspond to models with the metallicity evolutions of DA08+M08 and LN06, respectively, while other lines denote the results for fixed metallicity with $Z=0.02$ (dotted) and 0.004 (dashed).}
\label{fig:spm}
\end{figure*}

We compare the results for the flux of SRNs, ${\rm d}F(E_{\bar{\nu}_e})/{\rm d}E_{\bar{\nu}_e}$, and the event rate spectrum, ${\rm d}N_{e^+}(E_{e^+})/{\rm d}E_{e^+}$, for different metallicity evolution models in Figure~\ref{fig:spm}. Here, we also show the results for fixed metallicity with $Z=0.02$ and 0.004. The flux and event rate are higher for the case with $Z=0.004$ because they increase with the inclusion of failed supernovae \citep[][]{luna09,lien10,self13c}. In our model, the fraction of black-hole-forming progenitors for $Z=0.004$ is fixed to $\sim$0.14 while that for $Z=0.02$ is 0. The accretion in failed supernovae continues until the mass of the proto-neutron star reaches the maximum mass of neutron stars. Therefore, more potential energy can be released from black hole formation than for ordinary CCSNe leaving neutron stars. Furthermore, the mean energy of neutrinos emitted from failed supernovae is higher than that for an ordinary CCSN because the neutrino-emitting surface of the core continues to be heated. Thus, the difference in the spectra for $Z=0.02$ and 0.004 is larger in the high-energy regime.

\begin{figure*}
\plotone{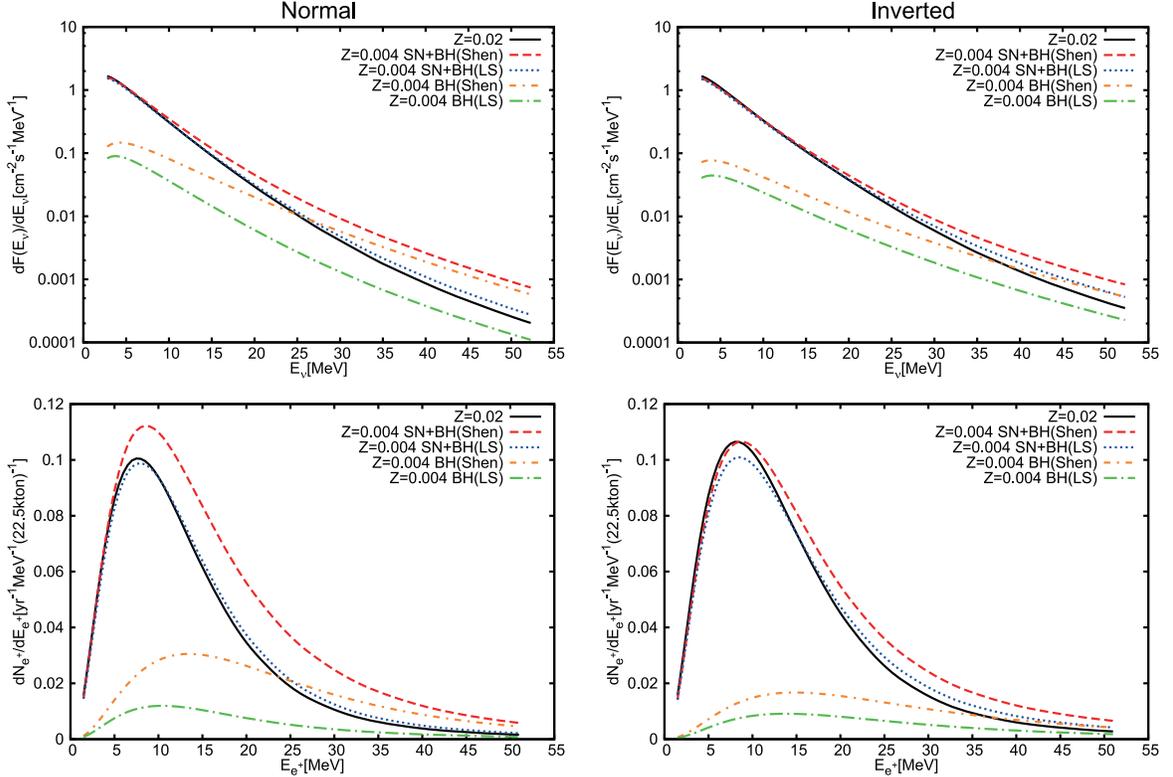}
\caption{Same as Figure~\ref{fig:spm} but for different EOSs. Dashed and dotted lines correspond to models with the Shen EOS and LS EOS, respectively, with the metallicity fixed to $Z=0.004$. The contribution of black-hole-forming failed supernovae is extracted for the Shen EOS (dot-short-dashed) and LS EOS (dot-long-dashed). Solid lines denote the results for fixed metallicity with $Z=0.02$, where failed supernovae are not included.}
\label{fig:spe}
\end{figure*}

The enhancement of SRNs due to failed supernovae is larger for the normal mass hierarchy than for the inverted mass hierarchy because neutrinos from the accretion of matter dominate in the case of failed supernovae. As already stated, $\nu_e$ and $\bar\nu_e$ are emitted more abundantly than $\nu_x$ ($=\nu_\mu=\bar \nu_\mu=\nu_\tau=\bar \nu_\tau$) from the accretion of matter. Nevertheless, for the inverted mass hierarchy, all $\bar{\nu}_e$ are converted to $\bar \nu_\mu$ or $\bar \nu_\tau$ as in Equation~(\ref{eq:neuoscih}). Meanwhile, for the normal mass hierarchy, the survival probability of $\bar{\nu}_e$ is 0.68 as in Equation~(\ref{eq:neuoscnh}). On the other hand, the neutrino spectra of ordinary supernovae do not depend strongly on flavor because neutrinos of all species are emitted equivalently during the cooling of the proto-neutron star. Therefore, we can see the contribution of failed supernovae more clearly for the normal mass hierarchy.

Models with the metallicity distribution functions of DA08+M08 and LN06 reside between the two fixed-metallicity cases with $Z=0.02$ and 0.004 (Figure~\ref{fig:spm}). The spectra of DA08+M08 and LN06 are similar. As shown in Table~\ref{tab:enme}, this is also the case for the event rates in various ranges of positron energy. In previous studies \citep[][]{yuk12,self13c}, the metallicity evolution of galaxies was not considered in the redshift dependence of the black hole formation rate. However, the black hole formation rate can be assessed using suitable metallicity evolution models of galaxies. In fact, in our model, the uncertainty of the black hole formation rate in SRNs is small for the given mass and metallicity ranges of black-hole-forming progenitors. On the other hand, as already mentioned, the mass and metallicity dependences of the fate of progenitors are still unclear. For the normal mass hierarchy, the event rate in the positron energy range of 10-26~MeV for the model with a fixed $Z$ of 0.004 is $\sim$40\% larger than that when $Z=0.02$. Thus, when the fraction of black-hole-forming progenitors is $\sim$0.3 \citep[][]{hori14}, the event rate would become $\sim$80\% larger than that for the model with $Z$ fixed to 0.02. In contrast, for the inverted mass hierarchy, the enhancement of the event rate is $\sim$15\% for $Z=0.004$ and, therefore, it would be $\sim$30\% when the fraction of black-hole-forming progenitors is $\sim$0.3.

Now we move on to the EOS dependence of black hole formation while the above investigations in this subsection were based on the models with the Shen EOS. In Figure~\ref{fig:spe}, we show the results obtained with the Shen EOS and LS EOS for fixed metallicity with $Z=0.004$ for the purpose of comparison. We can see that, as anticipated from \S~\ref{sec:bh}, the flux and event rate are lower for the LS EOS. In general, a stiffer EOS can result in higher SRN flux, which is consistent with a previous study \citep[][]{luna09}. In particular, for the LS EOS, the flux of the model with $Z=0.004$ does not significantly differ from that with $Z=0.02$, which does not include failed supernovae. The contribution of failed supernovae is also not clear for the event rates in various ranges of positron energy (Table~\ref{tab:enme}). In conclusion, the enhancement of SRNs due to failed supernovae is significant for the case with the normal mass hierarchy and a stiff EOS.

\begin{deluxetable*}{ccccccccc}
\tabletypesize{\footnotesize}
\tablewidth{0pt}
\tablecaption{SRN event rates in various ranges of positron energy in Super-Kamiokande over 1 year (i.e., per 22.5~kton~year) for models with CSFRD of DA08 and shock revival time of $t_{\rm revive}=200$~ms.}
\tablehead{ &  & \multicolumn{3}{c}{Normal mass hierarchy} & \ & \multicolumn{3}{c}{Inverted mass hierarchy} \\ \cline{3-5} \cline{7-9}
 metallicity evolution & EOS for BH & 18-26 & 10-18 & 10-26~MeV & & 18-26 & 10-18 & 10-26~MeV}
\startdata
$Z=0.02$       &         & 0.227 & 0.549 & 0.776 & & 0.301 & 0.640 & 0.941 \\
 DA08$+$M08 & Shen & 0.274 & 0.604 & 0.879 & & 0.326 & 0.660 & 0.986 \\
 LN06             & Shen & 0.288 & 0.625 & 0.912 & & 0.334 & 0.669 & 1.003 \\
 $Z=0.004$     & Shen & 0.387 & 0.714 & 1.100 & & 0.378 & 0.691 & 1.069 \\ \hline
 $Z=0.02$       &         & 0.227 & 0.549 & 0.776 & & 0.301 & 0.640 & 0.941 \\
 DA08$+$M08 & LS    & 0.233 & 0.554 & 0.787 & & 0.308 & 0.640 & 0.948 \\
 LN06             & LS    & 0.235 & 0.557 & 0.791 & & 0.311 & 0.642 & 0.953 \\
 $Z=0.004$     & LS    & 0.247 & 0.563 & 0.810 & & 0.322 & 0.632 & 0.954
\enddata
\label{tab:enme}
\end{deluxetable*}

\subsection{Dependence of Cosmic Star Formation Rate Density}\label{sec:dcsfrd}

\begin{figure*}
\plotone{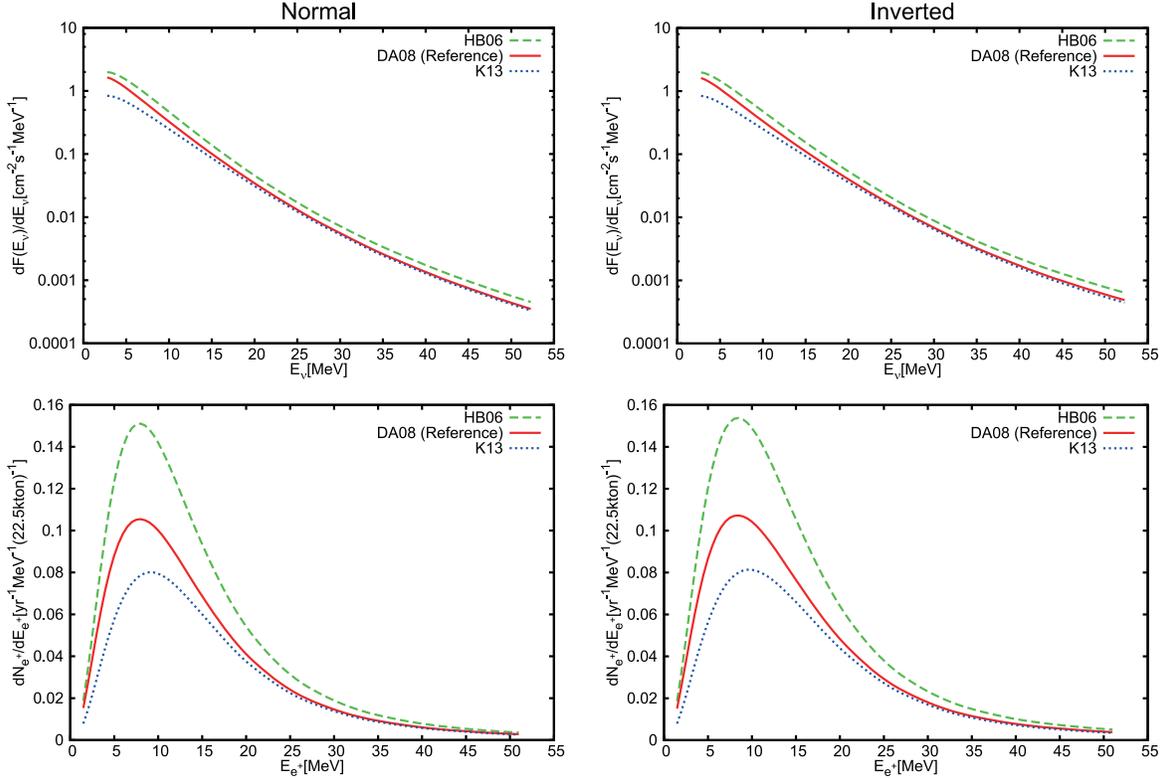}
\caption{Same as Figure~\ref{fig:spm} but for the models with metallicity evolution of DA08+M08, shock revival time of $t_{\rm revive}=200$~ms and Shen EOS. Dashed, solid and dotted lines correspond to models with the CSFRD of HB06, DA08 and K13, respectively.}
\label{fig:spsfr}
\end{figure*}

\begin{figure*}
\plotone{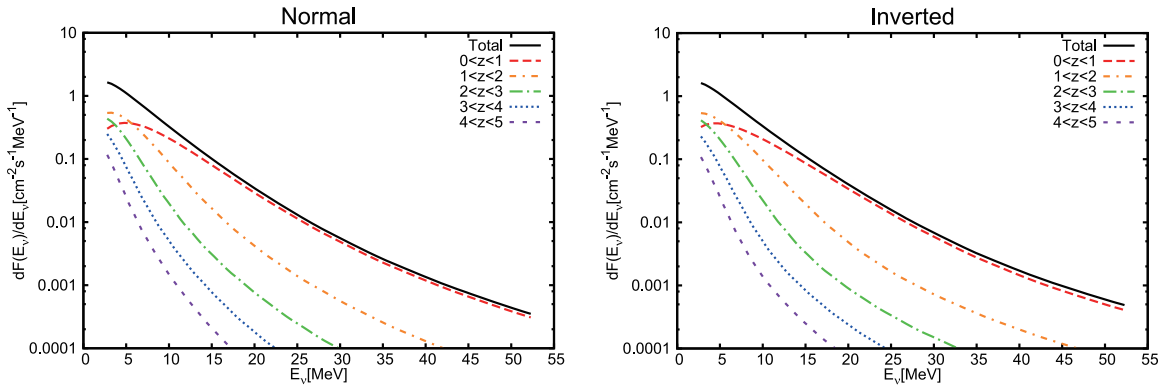}
\caption{Total fluxes of SRNs (solid) and contributions from various redshift ranges for the reference model. The lines except for the solid line correspond, from top to bottom, to the redshift ranges $0<z<1$, $1<z<2$, $2<z<3$, $3<z<4$ and $4<z<5$, for $E_\nu > 10$~MeV. The left and right panels show the cases for normal and inverted mass hierarchies, respectively.}
\label{fig:sprs}
\end{figure*}

\begin{figure*}
\plotone{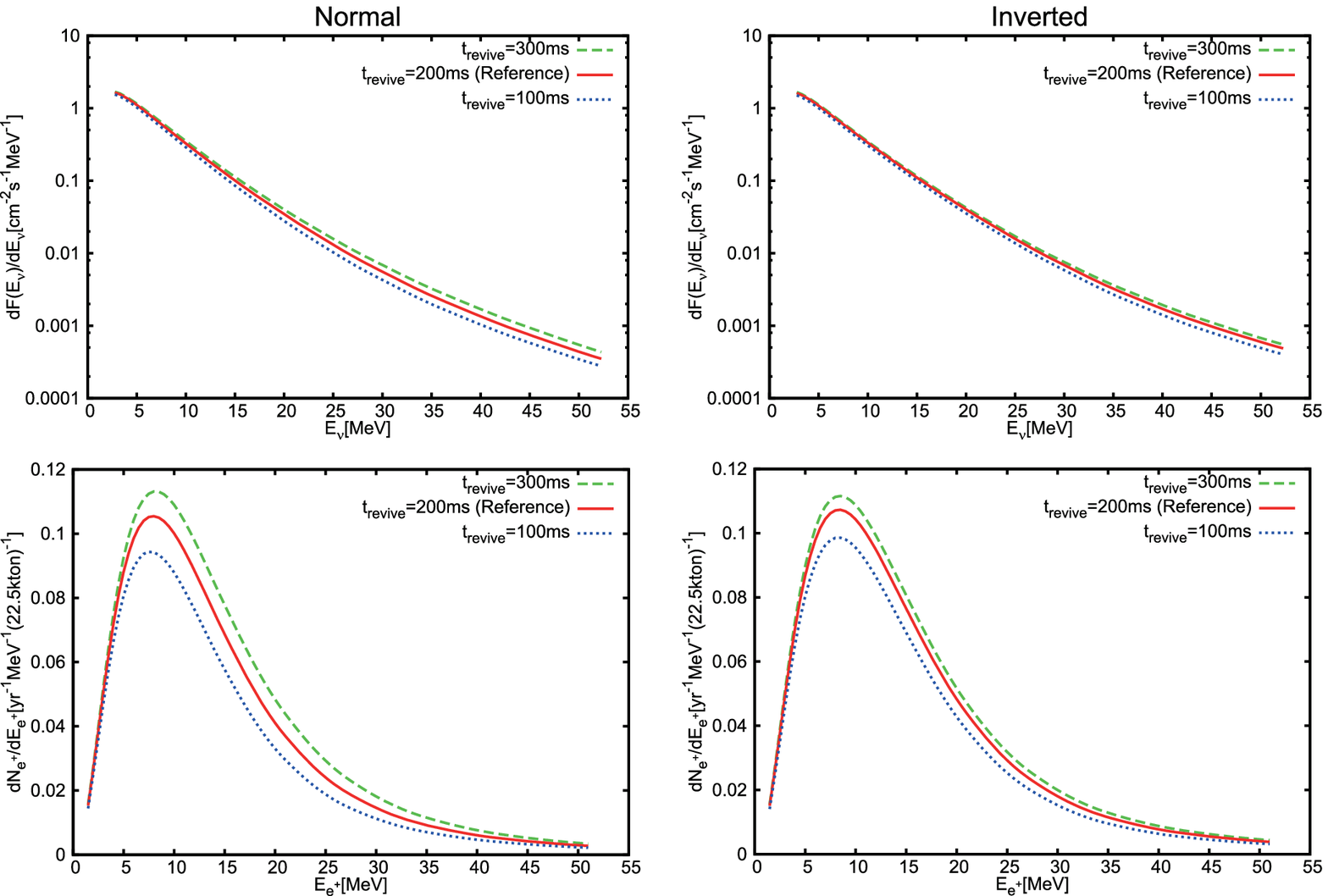}
\caption{Same as Figure~\ref{fig:spm} but for models with CSFRD of DA08 and metallicity evolution of DA08+M08 and Shen EOS. Dotted, solid and dashed lines correspond to models with shock revival times of $t_{\rm revive}=100$, 200 and 300~ms, respectively.}
\label{fig:spt}
\end{figure*}

The results for the SRN flux and event rate spectrum for various CSFRD models are shown in Figure~\ref{fig:spsfr}. A model with a higher CSFRD has more progenitors of both ordinary and failed supernovae as in Equation~(\ref{eq:ccrate}). We can recognize that the difference is clearer for the low-energy regime, in contrast to the models in Figure~\ref{fig:spm}. This is because, reflecting the fact that the uncertainty in CSRFD is larger for a higher redshift, the difference in CSFRD among the models adopted here is large for the redshifts with $z \gtrsim 0.5$. Neutrinos from the distant Universe accumulate in the low-energy region of the SRN spectrum because they are redshifted by the cosmic expansion and their energy is reduced by a factor of $(1+z)^{-1}$. For the SRN flux of the reference model, contributions from various redshift ranges are indicated in Figure~\ref{fig:sprs}. We can see that, as discussed in previous works \citep[e.g.,][]{ando04,mathe14}, the considerable flux of low-energy SRNs is attributed to the range $1<z<2$ while the dominant flux originates from the local Universe ($0<z<1$) for neutrino energy $E_\nu > 10$~MeV.

\subsection{Dependence of Shock Revival Time}\label{sec:dsrt}

In Figure~\ref{fig:spt}, we show the spectra obtained from the models with different values of the shock revival time, which is introduced as a parameter reflecting the still unknown explosion mechanism \citep[][]{self13a}. The models with a longer shock revival time have a higher event rate because, as already stated in \S~\ref{sec:sn}, the total emission number and energy of supernova neutrinos increase with the shock revival time. The increase in flux is clear for the high-energy regime; this trend is similar to that observed with the inclusion of failed supernovae (\S~\ref{sec:mebh} and Figure~\ref{fig:spm}). Nevertheless, as reported by \citet{self13c}, the hardening of the SRN spectrum is not significant because a considerable fraction of neutrinos is emitted after the shock revival, where a proto-neutron star is not heated and the mean energy of neutrinos gradually decreases. Note that, the increase in flux (Figure~\ref{fig:spt}) is larger for the normal mass hierarchy because $\nu_e$ and $\bar\nu_e$ are abundant in the accretion phase, which is longer in the case of later shock revival.

\subsection{Summary of Uncertainties}\label{sec:unct}

\begin{deluxetable*}{ccccccccccc}
\tabletypesize{\footnotesize}
\tablewidth{0pt}
\tablecaption{SRN event rates in various ranges of positron energy in Super-Kamiokande over 1 year (i.e., per 22.5~kton~year) for models with metallicity evolution of DA08+M08.}
\tablehead{  &  &  & \multicolumn{3}{c}{Normal mass hierarchy} & \ & \multicolumn{3}{c}{Inverted mass hierarchy} & \\ \cline{4-6} \cline{8-10}
CSFRD & $t_{\rm revive}$ & EOS for BH & 18-26 & 10-18 & 10-26~MeV & & 18-26 & 10-18 & 10-26~MeV & Figure~\ref{fig:spr}}
\startdata
HB06 & 100~ms & Shen & 0.286 & 0.704 & 0.990 & & 0.375 & 0.832 & 1.207 & \\
         &       & LS     & 0.227 & 0.635 & 0.863 & & 0.351 & 0.806 & 1.156 & \\
         & 200~ms & Shen & 0.361 & 0.833 & 1.193 & & 0.429 & 0.920 & 1.349 & \\
         &       & LS     & 0.302 & 0.764 & 1.066 & & 0.404 & 0.893 & 1.297 & \\
         & 300~ms & Shen & 0.432 & 0.938 & 1.370 & & 0.463 & 0.967 & 1.431 & Maximum \\
         &       & LS     & 0.374 & 0.869 & 1.242 & & 0.439 & 0.941 & 1.379 & \\
DA08 & 100~ms & Shen & 0.219 & 0.515 & 0.734 & & 0.286 & 0.598 & 0.885 & \\
         &       & LS     & 0.178 & 0.464 & 0.642 & & 0.269 & 0.578 & 0.847 & \\
         & 200~ms & Shen & 0.274 & 0.604 & 0.879 & & 0.326 & 0.660 & 0.986 & Reference \\
         &       & LS     & 0.233 & 0.554 & 0.787 & & 0.308 & 0.640 & 0.948 & \\
         & 300~ms & Shen & 0.326 & 0.677 & 1.003 & & 0.350 & 0.694 & 1.044 & \\
         &       & LS     & 0.285 & 0.627 & 0.911 & & 0.333 & 0.674 & 1.007 & \\
K13   & 100~ms & Shen & 0.203 & 0.443 & 0.645 & & 0.264 & 0.505 & 0.769 & \\
         &       & LS     & 0.171 & 0.410 & 0.581 & & 0.252 & 0.492 & 0.744 & Minimum \\
         & 200~ms & Shen & 0.252 & 0.514 & 0.767 & & 0.298 & 0.554 & 0.853 & \\
         &       & LS     & 0.221 & 0.482 & 0.703 & & 0.286 & 0.542 & 0.827 & \\
         & 300~ms & Shen & 0.298 & 0.570 & 0.868 & & 0.319 & 0.580 & 0.899 & \\
         &       & LS     & 0.266 & 0.537 & 0.804 & & 0.306 & 0.568 & 0.874 &  
\enddata
\label{tab:enste}
\end{deluxetable*}

\begin{figure*}
\plotone{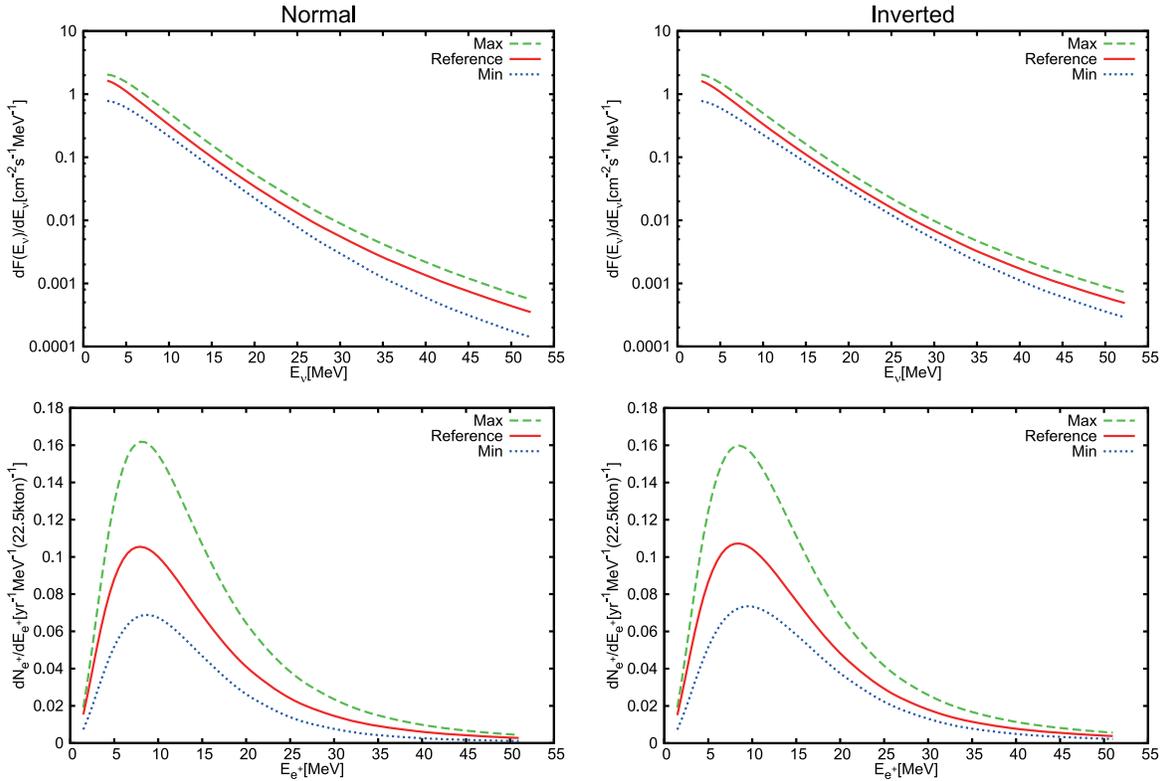}
\caption{Same as Figure~\ref{fig:spm} but for reference model (solid) and models with maximum (dashed) and minimum (dotted) values of SRN event rate among models with metallicity evolution of DA08+M08. See Table~\ref{tab:enste} for the parameter sets of the maximum and minimum models.}
\label{fig:spr}
\end{figure*}

\begin{figure}
\plotone{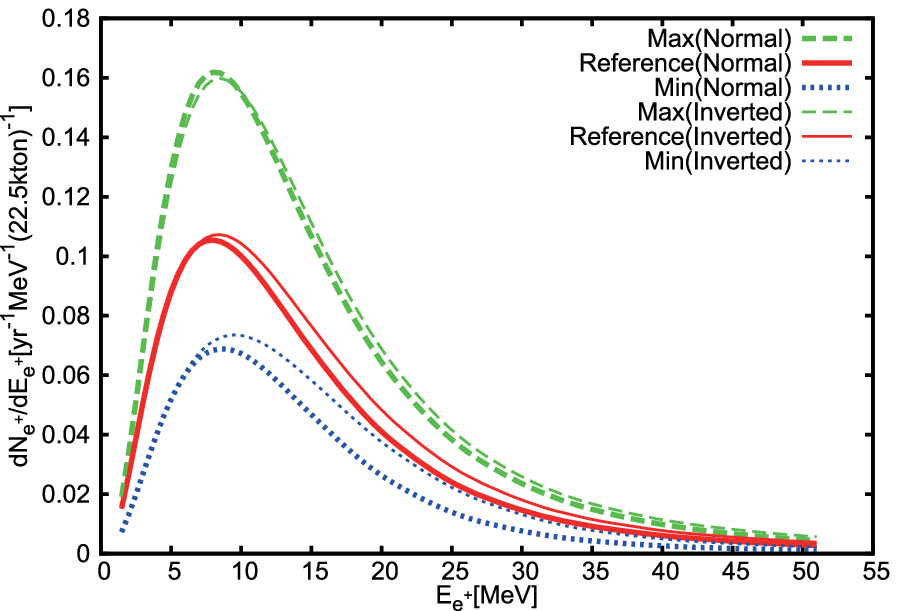}
\caption{Event rate spectra in Super-Kamiokande over 1 year obtained using reference model (solid) and models with maximum (dashed) and minimum (dotted) values of SRN event rate among models with metallicity evolution of DA08+M08. Thick and thin lines correspond to the results for the normal and inverted mass hierarchies, respectively.}
\label{fig:spr2}
\end{figure}

Here, we summarize the uncertainties in the SRN spectrum due to the EOS for black hole formation, CSFRD and shock revival time $t_{\rm revive}$, adopting the models with DA08+M08 for metallicity evolution, whose variance was shown to be small in \S~\ref{sec:mebh}. In Table~\ref{tab:enste}, the event rates in various ranges of positron energy are listed for all combinations. We can see that the uncertainty of CSFRD is relatively larger than those of the EOS and $t_{\rm revive}$ for events in a low-energy range (10~${\rm MeV} \le E_{e^+} \le 18$~MeV), whereas they are comparable for events in a high-energy range (18~${\rm MeV} \le E_{e^+} \le 26$~MeV). This is because the SRN flux dependences on the EOS and $t_{\rm revive}$ are minor for low energies. This tendency is especially clear in the case of the inverted mass hierarchy, where the dependences on the EOS and $t_{\rm revive}$ are again minor. In Figure~\ref{fig:spr}, we show the spectra of the reference model and models with the maximum and minimum values of the SRN event rate given in Table~\ref{tab:enste}. We can recognize that the uncertainty in the SRN flux is a factor of two or three across the full energy range shown in Figure~\ref{fig:spr}. As already described, the difference in the low-energy spectrum mainly originates from CSFRD (see Figure~\ref{fig:spsfr}), whereas the choice of EOS and $t_{\rm revive}$ also affects the high-energy spectrum (see Figures~\ref{fig:spe} and \ref{fig:spt}).

We compare the results for the normal and inverted mass hierarchies in Figure~\ref{fig:spr2}. The difference is clear for model with minimum values of SRN event rate, where a short shock revival time ($t_{\rm revive}=100$~ms) and LS EOS are adopted. In models with short shock revival time (see Figure~\ref{fig:snspc}) and LS EOS (see Figure~\ref{fig:bhspc}), the difference in the spectra for $\bar \nu_e$ and $\nu_x$ before the neutrino oscillation is larger in the high-energy regime. This is for the following reason. The emission of high-energy $\nu_e$ and $\bar \nu_e$ is suppressed owing to dense accreting matter in early stage of the accretion phase ($\lesssim$100~ms after the bounce). The average energy of $\nu_x$ is higher even in the early stage \citep[see Figure~14 of][]{self13a} because a mean free path of $\nu_x$ is longer than those of $\nu_e$ and $\bar \nu_e$. On the other hand, high-energy $\nu_e$ and $\bar \nu_e$ are more efficiently emitted than $\nu_x$ later on the accretion phase ($\gtrsim$200~ms after the bounce) and the difference in the total emission number of $\bar \nu_e$ and $\nu_x$ gets smaller. Thus the difference in the SRN spectra for the normal and inverted mass hierarchies is clear for models with a short mass accretion. Note that LS EOS model has a shorter time to black hole formation, which corresponds to the duration of mass accretion, than Shen EOS model.

\section{Conclusion and Discussion}\label{sec:concl}

In this paper, we have studied the supernova relic neutrino (SRN) spectrum and event rate involving black-hole-forming failed supernovae. To evaluate the contribution of failed supernovae to SRNs, cosmic metallicity evolution is important as well as the mass and metallicity dependences of the fate of progenitors. In this study, impact of metallicity evolution on SRNs has been investigated for the first time. We have determined the redshift evolution of the metallicity distribution function of progenitors from observed galaxy stellar mass function and empirical relation of galaxy mass and metallicity. As a result, the uncertainty of the black hole formation rate in our model spectra of SRNs is small for the given mass and metallicity ranges of black-hole-forming progenitors. Our reference model of SRNs has a lower flux compared with previous studies, where a higher cosmic star formation rate density (CSFRD) \citep[][]{hb06} and a higher mean energy of emitted neutrinos were assumed \citep[e.g.,][]{beacom10}.

We have also investigated the dependences of SRNs on the CSFRD, shock revival time $t_{\rm revive}$ and equation of state (EOS). The shock revival time is introduced as a parameter that should depend on the still unknown explosion mechanism of core collapse supernovae. The EOS dependence has been considered for failed supernovae, whose collapse dynamics and neutrino emission are certainly affected, using the new result on a failed supernova with a different EOS. It has been found that the differences with regard to $t_{\rm revive}$ and the EOS are clear for the high-energy regime, especially in the case of the normal mass hierarchy. In contrast, the low-energy spectrum of SRNs is mainly determined by CSFRD for both mass hierarchies. Therefore, the detection of low-energy neutrinos is mandatory so as to probe the cosmic star formation history by SRN observation.

Although numerous noise events prevent the observation at low energies, considerable effort is now being devoted to reducing them. Photomultipliers are used to observe Cherenkov photons (in a water Cherenkov type detector such as Super-Kamiokande) or scintillation photons (in a liquid-scintillator-type detector such as KamLAND) emitted from positrons produced in the inverse beta decay (\ref{eq:ibd}). Meanwhile, the neutrons produced in Equation~(\ref{eq:ibd}) are captured on protons and emit delayed gamma rays (2.2~MeV). Therefore, $\bar{\nu}_{\rm e}$ events are identified by detecting 2.2~MeV gamma rays in KamLAND and recently in Super-Kamiokande \citep[][]{sk4srn}. Distinguishing $\bar{\nu}_{\rm e}$ events from other background events without neutron emission such as invisible muon events is called neutron tagging. Furthermore, neutron tagging will be more efficient with gadolinium-loaded water because of the high neutron capture rate on gadolinium and the high-energy (8~MeV) gamma-ray emission \citep[][]{beacom04}. Actually, gadolinium loading is planned in Super-Kamiokande. This will allow the detection of low-energy events in the range of 10~${\rm MeV} \le E_{e^+} \le 18$~MeV investigated in our study as a regime where the event number is mainly determined by the CSFRD. In fact, assuming 90\% of the neutron-tagging efficiency \citep[][]{wata09}, Super-Kamiokande will detect 4-9 low-energy events over 10 years depending on CSFRD. We hope that Super-Kamiokande with gadolinium-loaded water will become a powerful detector of SRNs, opening a new door to exploring the history of the Universe.

\acknowledgments

The authors are grateful to Makoto Sakuda and Tomonori Totani for valuable comments. We would also like to thank Yusuke Koshio for providing us the data of upper limits from Super-Kamiokande. In this work, numerical computations were partially performed on the supercomputers at Research Center for Nuclear Physics (RCNP) in Osaka University. This work was partially supported by Grants-in-Aids for the Scientific Research (No.~24105008, No.~26104006, No.~26105515 and No.~26870615) from MEXT in Japan.

\begin{figure}[h]
\plotone{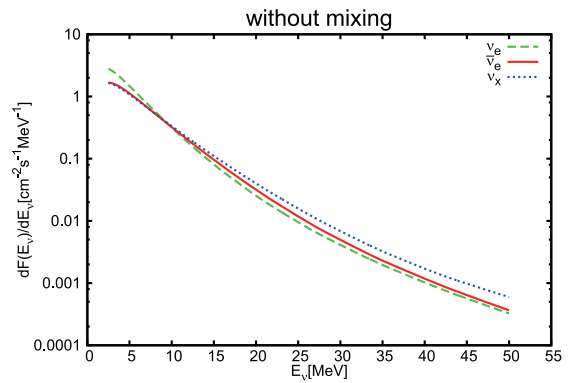}
\caption{Fluxes of SRNs obtained using reference model for the case without neutrino oscillation. Dashed, solid and dotted lines correspond to $\nu_e$, $\bar\nu_e$ and $\nu_x$ ($=\nu_\mu=\bar \nu_\mu=\nu_\tau=\bar \nu_\tau$), respectively.}
\label{fig:apdx}
\end{figure}

\appendix

\section{SRN Flux Data Table} \label{fluxtable}
For general use in any research for astronomy, astrophysics, and physics, the numerical data of SRN flux in this study are publicly available on the Web at\\
{\tt http://asphwww.ph.noda.tus.ac.jp/srn/}\\
We provide not only the reference model but also models with maximum and minimum values of SRN event rate among models with metallicity evolution of DA08+M08. Moreover, for a broad application, fluxes of SRNs with $\nu_{e}$, $\bar{\nu}_{e}$, and $\nu_{x}$ for the case without neutrino oscillation, which are shown in Figure~\ref{fig:apdx}, are also available.

\end{document}